# Flexible Entanglement Distribution Overlay for Cloud/Edge DC Interconnect as Seed for IT-Secure Primitives

**Fabian Laudenbach, Bernhard Schrenk, Martin Achleitner, Nemanja Vokić, Dinka Milovančev, and Hannes Hübel**

*AIT Austrian Institute of Technology, Center for Digital Safety&Security / Security & Communication Technologies, 1210 Vienna, Austria.*
*Author e-mail address: bernhard.schrenk@ait.ac.at*

We leverage spectral assets of entanglement and spatial switching to realize a flexible distribution map for cloud-to-edge and edge-to-edge quantum pipes that seed IT-secure primitives. Dynamic bandwidth allocation and co-existence with classical control are demonstrated.

## 1. Introduction

The mission of 5G networks to provide communications, computing, control and content delivery, has pushed datacenter (DC) resources to the edge of the optical infrastructure. While exhaustive processing in DC warehouses remains centralized in the cloud, micro-scale DCs are being now distributed closer to the end-user in order to provide real-time performance at much lower latency, thus enabling an elevated quality-of-experience [1]. However, as critical data and infrastructural assets move deeper into the field, which up to now has been mostly laid out as a passive and hence "dumb" last mile, security aspects move into the focus in order to retain a trustful and secure network. With quantum computing on the verge, the IT-secure primitives are needed more than ever. Quantum entanglement, one of the best known and perplexing hallmarks of quantum mechanics, can be a useful resource and an essential ingredient to ensure the highest level of security in such a hybrid cloud / edge datacenter infrastructure.

In this work, we present a practical entanglement distribution scheme that permits flexible distribution of quantum resources between any two edge sites but also between the cloud and the edge. On-demand allocation of entangled wavelength channels and dynamic bandwidth allocation among the channels is experimentally demonstrated. We further prove the robust integration of a classical control channel at the quantum layer.

## 2. Quantum-Seeded IT-Security in a Hybrid Cloud/Edge DC Infrastructure

Figure 1 presents a typical scenario in which micro-DCs at the edge are integrated with a passive optical wireline-wireless network infrastructure. In order to provide the highest degree of security in this hybrid DC architecture, a quantum-optic overlay serves the provision of quantum correlations between several sites. IT-secure primitives can then be distilled from these stronger-than-classical correlations. The overlay is rolled-out together with the classical telecom infrastructure and can take advantage of fiber abundance or means of spatial multiplexing.

A flexible distribution among quantum assets is preferred in order to virtually provide a pipe between the cloud DC (Emma, **E**) and an edge micro-DC site (Alice to Diana, **A**-**D**) but also between any pair of edge DCs. This is facilitated by means of entanglement distribution, for which centralized Einstein-Podolsky-Rosen (EPR) sources are employed. While an asymmetric EPR source [2] provides connectivity between the cloud and an edge DC, a symmetric EPR source [3] interconnects two edge DCs. Since a source of entanglement contains no active encoding, there is no information present. In contrary to QKD, it could be placed in a completely untrusted environment, and even be operated by an adversary. The typical dual-feeder architecture of resilient optical network feeder networks is used to feed a flexible node that serves the purpose of spectral and spatial switching of quantum links, which allows to address any connection between DC sites while also adding a degree of dynamic bandwidth allocation for the quantum overlay.

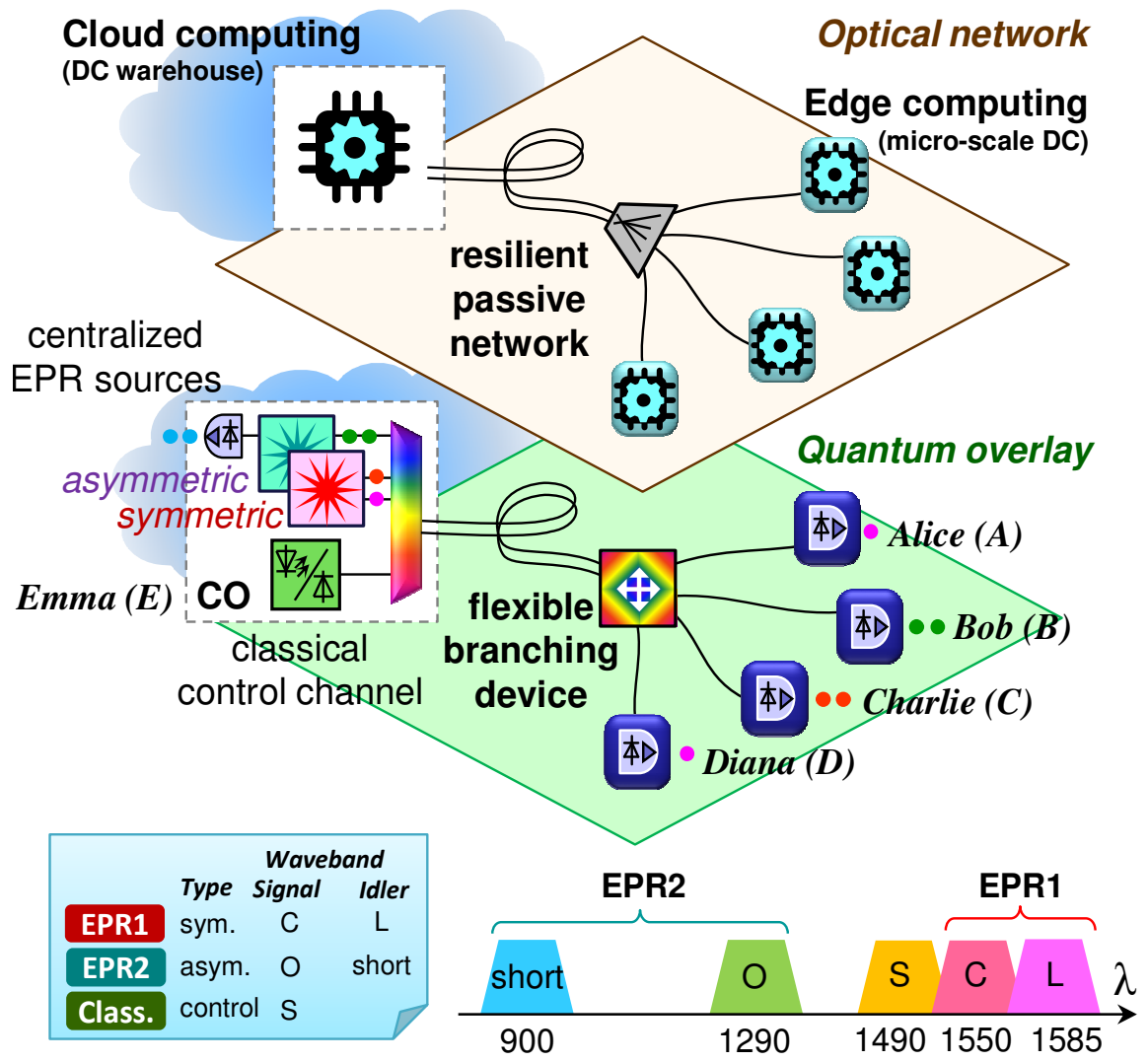

Fig. 1. Entanglement distribution for DC interconnect.

The flexible distribution is based on spectral slicing and spatial routing of entanglement within the network node. Two distribution mechanisms are provided for the four edge DC sites: The symmetric EPR1 source at **E**mma emits both photons in the C/L-band and enables, through the reconfigurable branching node, various mappings to edge sites **A**-**D**. In particular, EPR1 (Fig. 2b) produces two streams of entangled photon pairs, each consisting of a C/L-band combination. In order to allow the cloud DC (**E**) to share entanglement between the edge DCs, the asymmetric EPR2 source provides photons pairs with 900 / 1290 nm allocation. This allows efficient photo-detection of the short wavelength using silicon photodetectors while the O-band photon is transmitted to sites **A**-**D**.

Six exemplary distribution scenarios have been chosen (Fig. 2a): In mapping I, the edge DCs share entanglement from EPR1 (**A**-**B**: C-band, **C**-**D**: L-band), while there is no cloud-related activity. Mapping II is another edge-centric setting with reversed C/L-band allocation. Mapping III gives DC **A**lice a special role as she shares entanglement with two other DCs (**C**, **D**). This entanglement concentration elevates her rate as she receives both, L- and C-band photons. Moreover, edge DC **B**ob enjoys a high-rate quantum pipe of EPR2 with the cloud DC (**E**). In mapping IV, edge DCs **A** and **C** exploit both C/L wavelengths among them, while **D** is connected to the cloud (**E**). Mapping V shows a different site-wavelength allocation of I, while mapping VI resembles a concentration setting as in III.

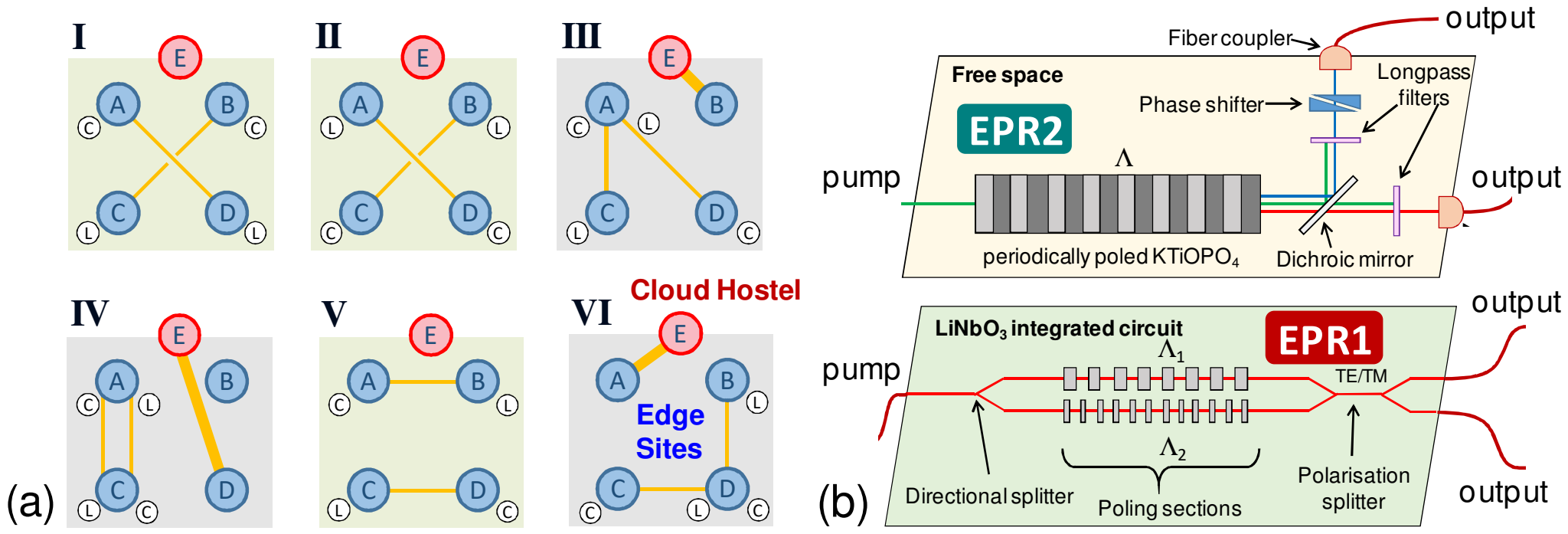

Fig. 2. (a) Distribution maps for edge-centric (I, II, V) distribution, entanglement concentration (III, VI) and cloud-to-edge distribution (III, IV, VI). (b) Employed EPR sources.

## 3. Experimental Setup for Entanglement Overlay at the DC Interconnect and Reconfigurable Network Node

Figure 3a shows the experimental setup. An auxiliary management and control (AMC) channel at 1490 nm is hosted together with the two EPR sources at the cloud hostel (**E**). At the reconfigurable network node the AMC channel is dropped out to instruct an 8×8 space switch that is used in a fold-back configuration to slice and re-assemble the EPR spectrum of the sources through waveband C/L splitters. The synthesized spectra for edge-to-edge interconnect and the exclusive O-band channel for cloud-to-edge interconnect are then forwarded to the edge sites (**A**-**D**). Figure 3b shows an exemplary distribution map as transmission *T* for **A**lice' node interface. Quantum signals in all three wavebands is realized at low fiber-to-fiber loss of 2.2 to 3.8 dB, depending on the actual loop-back configuration.

The ITU-T G.652B-compatible single-mode feeder and drop fiber spans between cloud hostel, node and edge sites are 12.8 and 4.3 km, respectively. The typically found resilient dual feeder is exploited to jointly feed the three EPR bands (2×C/L + O) and the S-band AMC channel. In order to avoid Raman scattering during co-propagation of AMC and quantum signals [4], classical control is temporally multiplexed and only active during reconfiguration.

The performance of entanglement distribution is evaluated through reception of the quantum signals through a pair of InGaAs single-photon detectors (SPADs) with one free-running and one gated SPAD, resulting in a detection efficiency of ~1%. The detector events are registered by a time-to-digital converter (TTM) for subsequent analysis of the coincidence rates between the various site pairs. A state analysis was then performed to evaluate the performance of entanglement distribution, for which polarization analyzer units have been included.

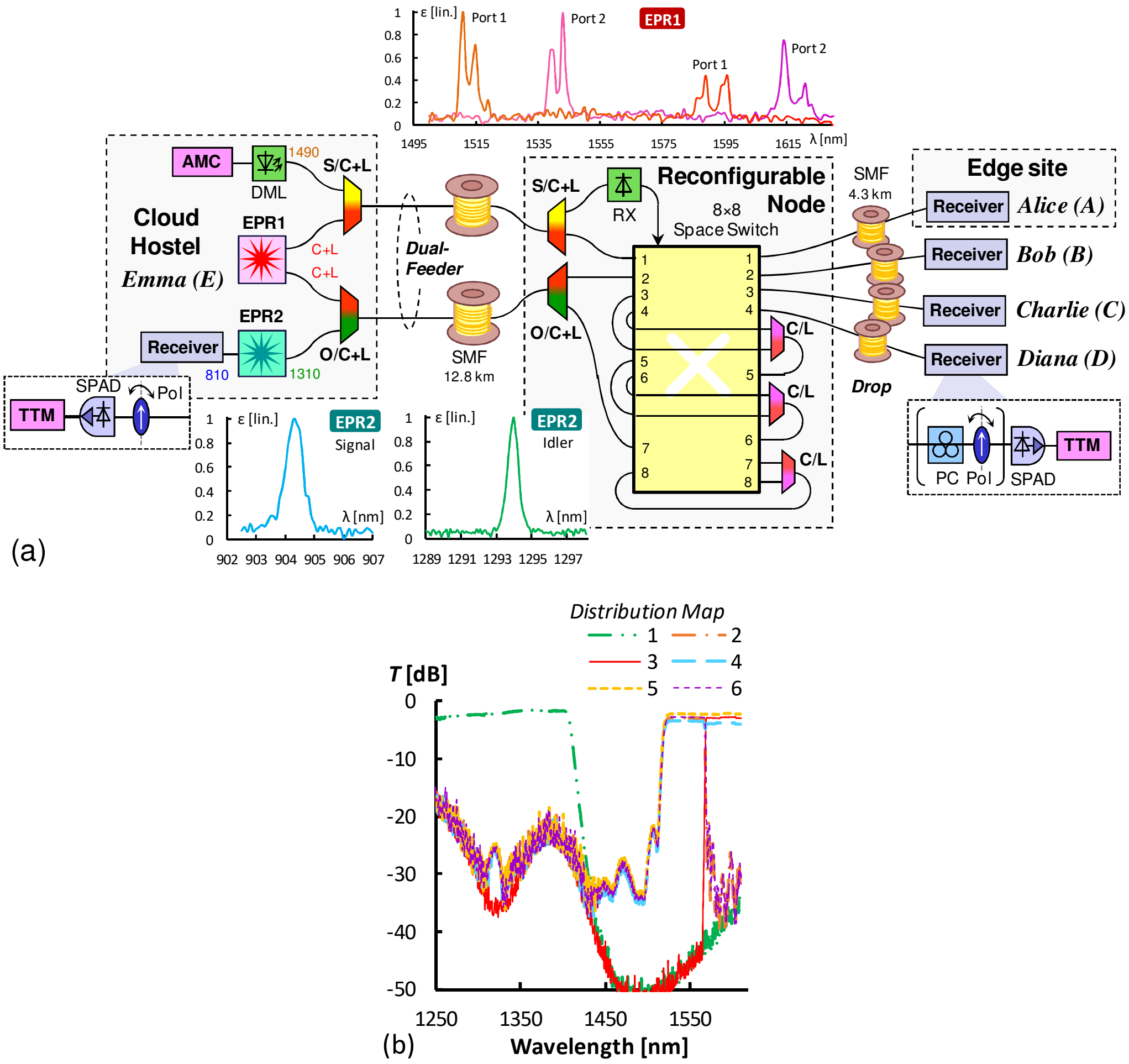

Fig. 3. (a) Experimental setup. The insets show the normalized emission spectra of the EPR sources.
(b) Spectral switching at Alice' node port.

## 4. Entanglement Distribution between any Cloud and Edge DCs, and Dynamic Bandwidth Allocation

The performance was first evaluated for various mappings in static operation in terms of pair-wise distribution of single photons and, subsequently, for entangled photons distribution. Finally, dynamic switching is validated.

For the distribution of photon pairs between edge sites, sourced by EPR1, the respective average coincidence rate varies between 2 and 6 coincidence counts/second (cc/s), which can be exploited as seed to periodically cycling a key for data encryption. There is a clear asymmetry between the two streams of entangled pairs, very notable in mappings I and II and III (Table I). The difference in the measured rate is however consistent with the mapping of the two streams (a certain C/L-band pair has lower rate) and we ascribe the difference to a lower generation amplitude as seen in the spectral distribution for EPR1 (see inset in Fig. 3a). Entanglement concentration is successfully achieved in mappings II and IV, as indicated by the higher rate for the users **A** and **D**, respectively. An increased bandwidth of >16 cc/s between users **A** and **C** is observed in mapping IV (Fig. 4a), for which the node is only traversed once, meaning a lower transmission loss. Higher rates can be obtained using two free-running detectors at a higher resulting detection efficiency, which has a square-dependence on the pair coincidence rate.

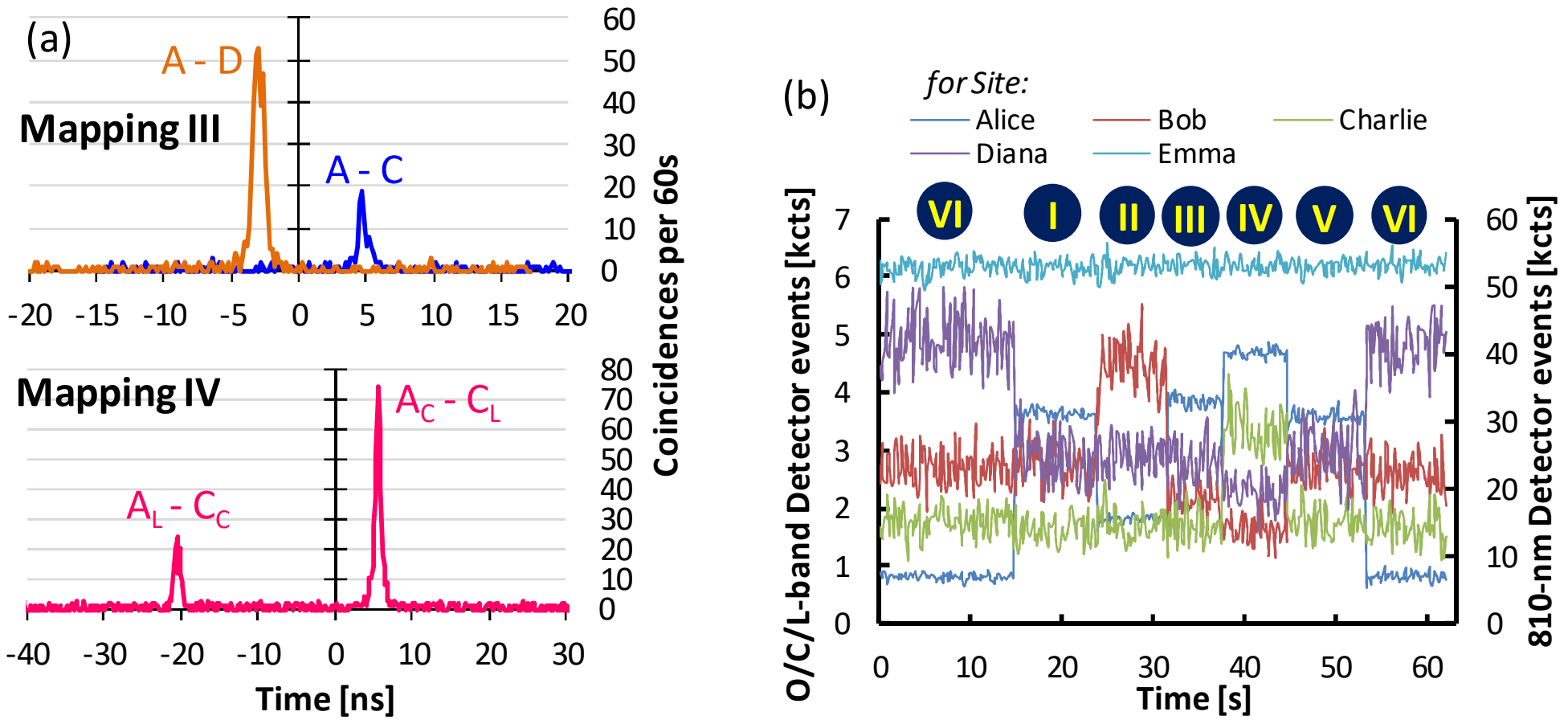

Fig. 4. (a) Coincidence rates among edge DCs. (b) Dynamic cycling through the distribution map.

Polarization entanglement as shared out by EPR2 between **E** and the edge sites **A-D** was analyzed through the detected coincidence rate. A back-to-back visibility measurement yielded V = 0.955 ± 0.033 and serves as reference for entanglement distribution. For the three mappings III, IV and VI, where entanglement is shared with **E**mma, the measured visibility drops to about 0.84 to 0.89 due to the low count rate, which complicates the polarization drift recovery along the transmission fiber. Nevertheless, all visibility measurements clearly exceeded the classical limit of $2\sqrt{2} \approx 0.707$, and thus fulfill the deployment needs of secure-IT primitives collapsed on top of the entanglement distribution. The local-to-remote (**E-A/D**) brightness amounted to 29.3 ± 2.3 cc/s for polarization-entangled photons.

Dynamic switching has been finally conducted through instructing the node to cycle through the entire distribution map (Fig. 2). The registered count rates at all edge sites are reported in Fig. 4b. The delivered photon rate can be dynamically adjusted by 4 dB among the edge sites (**A-D**), while the cloud hostel (**E**) retains a steady rate. The fast response of the space switch does not cause a gap larger than the 50-ms resolution for the acquisition of the delivered photon rates, yielding a quasi-hitless distribution. More importantly, there is no detrimental impact observed for co-propagation of the AMC channel in the S-band when compared to a purely electrically implemented AMC channel. This is explained by the short (1.6 ms) transmission for a node instruction in combination with the associated temporal gating of the AMC channel and its spectral separation from the quantum channels. This proves the feasibility of integrating a co-existing classical control channel at the quantum overlay.

| Map | Source and served sites | Coincid. rate [$s^{-1}$] |
|---|---|---|
| **I** | EPR1: A-D, B-C | 5.9 / 1.2 |
| **II** | EPR1: A-D, B-C | 1.4 / 5.9 |
| **III** | EPR1: A-C, A-D | 1.2 / 4.7 |
| | EPR2: E-B | N/A (V = 0.841) |
| **IV** | EPR1: 2× A-C | 8.8 / 7.8 |
| | EPR2: E-D | 29.3 (V = 0.886) |
| **V** | EPR1: A-B, C-D | 5.8 / 2.3 |
| | EPR2: E-A | N/A (V = 0.842) |
| **VI** | EPR1: B-D, C-D | 2.6 / 5.8 |

Table 1. Distribution performance.

## 5. Conclusions

We have experimentally demonstrated the flexible provisioning of single and entangled photons as security-enabling quantum overlay in a hybrid cloud/edge DC architecture. Dynamic allocation of quantum bandwidth between edge sites in a 4-dB range has been accomplished through reconfigurable slicing and tailoring from the O- to the L-band, and agile space switching, for which the co-existence with a classical control channel has been validated. Both, cloud-to-edge and edge-to-edge connectivity, have been demonstrated in six representative distribution maps. The EPR sources could be placed in the hand of an adversary without compromising the security in the DC interconnect.

## 6. Acknowledgement

This work has received funding from the EU Horizon-2020 R&I programme (grant agreement No 820474).